\documentclass[%
 aip,
 amsmath,amssymb,
preprint,%
reprint, onecolumn 
]{revtex4-1}

\usepackage{dcolumn}
\usepackage{mathptmx}
\usepackage{etoolbox}
\usepackage[english]{babel}
\usepackage[utf8]{inputenc}
\usepackage[T1]{fontenc}
\usepackage{amsmath}
\usepackage{amsfonts}
\usepackage{amssymb}
\usepackage{txfonts}
\usepackage{bm}
\usepackage{graphicx}
\usepackage{xcolor}
\usepackage{subfig}
\usepackage{cancel}
\usepackage{epstopdf, epsfig}
\usepackage{url}

\graphicspath{ {./obrazki/} }
\newcommand{\dd}{\textrm{d}}

\newcommand{\ddbarr}{\hspace{0.1em} \textrm{d} \hspace*{-0.15em}\bar{}\hspace*{0.02em} \hspace{0.1em}}

\makeatletter
\def\@email#1#2{%
 \endgroup
 \patchcmd{\titleblock@produce}
  {\frontmatter@RRAPformat}
  {\frontmatter@RRAPformat{\produce@RRAP{*#1\href{mailto:#2}{#2}}}\frontmatter@RRAPformat}
  {}{}
}%
\makeatother
\begin{document}

\preprint{}

\title[Thermodynamics of Rayleigh-B\'enard convection]{Thermodynamic selection of stationary Rayleigh-B\'enard flow states with slip and no-slip boundaries}

\author{Robert Ho\l yst}
 \thanks{These authors contributed equally to this work.}
 \email{rholyst@ichf.edu.pl, pzuk@ichf.edu.pl}
\affiliation{Institute of Physical Chemistry, Polish Academy of Sciences, Kasprzaka 44/52, 01-224 Warszawa, Poland}%
\author{Pawe\l{} J. \.Zuk}
 \thanks{These authors contributed equally to this work.}
\affiliation{Institute of Physical Chemistry, Polish Academy of Sciences, Kasprzaka 44/52, 01-224 Warszawa, Poland}%
\author{Konrad Gi\.zy\'nski}
\affiliation{Institute of Physical Chemistry, Polish Academy of Sciences, Kasprzaka 44/52, 01-224 Warszawa, Poland}%
\author{Jakub Wr\'obel}
\affiliation{Institute of Physical Chemistry, Polish Academy of Sciences, Kasprzaka 44/52, 01-224 Warszawa, Poland}%
\author{Anna Macio\l{}ek}
\affiliation{Max-Planck-Institut f\"ur Intelligente Systeme Stuttgart, Heisenbergstr. 3, D-70569 Stuttgart, Germany}
\affiliation{Institute of Physical Chemistry, Polish Academy of Sciences, Kasprzaka 44/52, 01-224 Warszawa, Poland}%

\author{Peter V. E. McClintock}
\affiliation{Department of Physics, Lancaster University, Lancaster LA1 4YB, UK}

\date{\today}

\begin{abstract}
Multiple stationary flow states may exist under the same external/boundary conditions, while the hydrodynamic equations alone do not provide a global criterion for determining which of these states is preferentially selected. Here, we formulate and test such a criterion for stationary Rayleigh–B\'enard (RB) convection within a global non-equilibrium thermodynamic description (Ho\l yst et al., EPL 149, 30001 (2025)). We find that each stationary flow state in RB convection is characterized by the state function $\mathcal{F}=-E_k-\Psi$,
where $E_k$ is the macroscopic kinetic energy of the flow and $\Psi$ is a thermodynamic state function that emerges from the nonlinear inertial term in the Navier–Stokes equation and is associated with the compression and redistribution of the gas accompanying the formation of the convective state. Changes in $\mathcal{F}$ correspond to path-independent work between stationary states. We show that, among dynamically accessible stationary states, the state with the lower value of $\mathcal{F}$ is more stable and is preferentially selected. Both the macroscopic kinetic energy acquired by the flow and the accompanying thermodynamic reorganization of the gas therefore contribute in the same direction to lowering $\mathcal{F}$ and stabilizing the stationary state. We test the criterion using numerical solutions of the hydrodynamic equations for Rayleigh–B\'enard convection with both slip and no-slip boundary conditions. For both types of boundary, the observed selection of competing stationary states is consistent with the thermodynamic prediction. The resulting state function provides a global criterion for the selection and stability of stationary hydrodynamic states. It establishes a direct connection between the nonlinear dynamics of the Navier–Stokes equation, thermodynamic work, and macroscopic flow stability.
\end{abstract}

\maketitle

\section{Introduction}

\subsection{Global Non-Equilibrium Thermodynamics of Stationary States}
 
A central aim of global non-equilibrium thermodynamics is to identify a small set of global state variables that gives a thermodynamically consistent account of energy changes in a spatially non-uniform system. The starting point is the local description provided by irreversible thermodynamics and fluid mechanics. Conservation of mass, momentum and energy determines the fields $\rho(\mathbf r,t)$, $\mathbf v(\mathbf r,t)$ and $T(\mathbf r,t)$, which generally vary in space and time \citep{deGroot2013}. Global thermodynamics, by contrast, describes changes in a system's internal energy, $\Delta E_T$, through the net heat $Q$, work $W$ and matter exchange $Z$:
\begin{equation} \label{eq:consEq}
  \Delta E_T=Q+W+Z.
\end{equation}
Its variables, such as entropy $S$, volume $V$ and mole number $N$, characterize the system as a whole. In (\ref{eq:consEq}), the exchange terms are positive when energy enters the system.

Our approach connects these two descriptions by integrating the local balance equations over space and time. In previously studied systems, this procedure has yielded global energy relations that retain the form of equilibrium thermodynamics, even though their variables describe non-uniform states \citep{holyst2025global,gizynski2025}. For example, under infinitesimal changes of the boundary conditions, the internal energy of an ideal gas in Poiseuille flow and heat flow obeys
\begin{equation}
  \mathrm d E_T=T^*\,\mathrm d S^*-p^*\,\mathrm d V+\mu^*\,\mathrm d N,
\end{equation}
where $T^*$, $p^*$, $S^*$ and $\mu^*$ are effective global quantities. They are defined from the non-uniform fields and depend on the system and its boundary conditions. For an ideal gas, $p^*=2E_T/(3V)$ and $T^*=2E_T/(3NR)$; the chemical potential $\mu^*$ has the ideal-gas form evaluated at $T^*$ and $p^*$. The effective entropy $S^*$ also has the equilibrium functional form, although it is not the volume integral of the local entropy: it excludes the dissipative background that sustains the flow and produces entropy \citep{holyst2025global}.

These effective quantities are not introduced as simple averages for their own sake. They provide a global thermodynamic description in which changes of the effective variables account consistently for changes in the system's energy. In general, the non-uniform system is mapped onto an effective uniform one. For example, the volume-averaged pressure and internal energy density define $p^*=\langle p(\mathbf r)\rangle$ and $E_T/V=\langle \rho(\mathbf r)u(\mathbf r)\rangle$, while the global temperature has been defined as $T^*=\langle\rho(\mathbf r)T(\mathbf r)\rangle/\langle\rho(\mathbf r)\rangle$. The precise definitions are system-dependent and are determined by the governing equations and boundary conditions.

This framework has been applied to ideal gases, binary mixtures, van der Waals fluids, gas columns under gravity, Couette and Poiseuille flows, photoreactors and continuously stirred tank reactors \citep{holyst2022thermodynamics,maciolek2023parameters,Holyst2023fundamental,Holyst2023gravity,makuch2023steady,holyst2025global,holyst2025cej}. These studies motivate a generalized second law for transitions between stationary states \citep{holyst2024direction, holyst2025global}. Its meaning is clearest when the exchanges during a transition are stated explicitly.

Consider two stationary states under fixed external constraints. For the closed RB cell studied here, there is no matter exchange. Let $Q$ denote the net heat entering the system during the transition, and $W$ the net work done on it, with both quantities referring to the transition rather than to the continuous heat and work needed to maintain the stationary state. The energy balance can then be written
\begin{equation}
  \Delta E_T-Q-W=W_{\mathrm{ext}},
\end{equation}
where $W_{\mathrm{ext}}$ is the work done on the system by an auxiliary external device; this equation fixes the sign convention. The generalized second law states that a spontaneous transition must satisfy
\begin{equation} \label{eq:inEqWext}
  W_{\mathrm{ext}}\leq 0.
\end{equation}
Thus, in a spontaneous transition the system can deliver net work to the device, whereas the device cannot deliver net work to the system. Equality applies in the reversible limit.

This statement is a non-equilibrium extension of the maximum-work principle: for specified external constraints and exchanges, the work that can be extracted from a transition is bounded by the thermodynamic change between its initial and final states. To apply the principle to stationary states with the same boundary conditions, we consider a quasi-static path made up of successive stationary states. Such a path is generally unavailable through a change of the boundary parameters alone. The auxiliary device represents a controlled actuator and work reservoir that makes the path possible-for example, by applying an external force during the transition-and accounts for the work exchanged along it. It is a device for defining and measuring the work of the constrained path, not a source of unaccounted energy. ``Quasi-static'' here means that the path passes through stationary states; it does not mean that the fluid velocity in those states is zero. The local flow may be finite, while its fields remain time-independent.

Within this framework, a state is more stable in the global thermodynamic sense if it is preferentially selected in a spontaneous transition under the specified constraints. This criterion concerns selection between competing stationary states; it is distinct from local dynamical stability to small perturbations. The inequality (\ref{eq:inEqWext}), together with the energy balance, supplies the basis for comparing such states. In the present work, we use it to determine whether the global thermodynamic description can also account for the energy associated with macroscopic fluid motion.

A pedagogical introduction to non-equilibrium thermodynamics, presented as a lecture, is available online \citep{robertWWW}.

\subsection{Rayleigh-B\'enard Convection}

Rayleigh-B\'enard (RB) convection occurs when a fluid is heated from below and cooled from above in a gravitational field. When the imposed temperature difference is sufficiently large, buoyancy can drive a stationary convective flow. The phenomenon has been studied since the nineteenth century \citep{oberbeck1879warmeleitung, lorenz1881leitungsvermogen, thomson1882changing, benard1900tourbillons, boussinesq1903theorie, rayleigh1916lix} and remains an active subject of experimental and theoretical research \citep{cross1993pattern, Bodenschatz2000, Ahlers2009, ecke2023turbulent}.

Much of the RB literature focuses on the onset, structure and dynamics of convection. Thermodynamics has most often entered through local entropy transport and production, alongside analyses of energy budgets, heat transfer and viscous dissipation \citep{shang2005test, kerr2001energy, hughes2013available,siggia1994high, urban2011efficiency,shishkina2007local, Ahlers2009, lohse2010small}. These approaches provide important information about local processes and sustained fluxes. The global thermodynamic question considered here is different: how does the energy stored in a stationary state change when the system moves between competing flow configurations?

The distinction matters because total entropy or entropy production need not, by itself, provide a useful global measure for selecting stationary states far from equilibrium \citep{Holyst2023fundamental, holyst2025global}. Entropy profiles remain valuable for understanding local behavior \citep{mishra2010energy}, and entropy-based selection principles have also been proposed \citep{kita2006entropy, kita2006principle,ban2020thermodynamic}. Here we instead track the net heat and work associated with a transition between stationary states and ask whether these global energy changes define a thermodynamic preference.

In this work global non-equilibrium thermodynamics is extended to RB convection by including macroscopic kinetic energy and the additional energetic contribution associated with the nonlinear inertial term in the Navier--Stokes equation. The latter reflects the thermodynamic reorganization of a compressible fluid as its stationary flow changes; it is not captured by kinetic energy alone. We derive the relevant energy balance and evaluate the net heat and work for transitions between stationary states. This leads to a state function whose value can be compared across competing states: among dynamically accessible stationary solutions under the same constraints, the lower value identifies the thermodynamically preferred state. This prediction is tested using numerical solutions for RB convection with both slip and no-slip boundary conditions.

The aim is therefore not only to characterize individual convective solutions, but to establish a global criterion for thermodynamic selection among them. By comparing the criterion with the stationary states obtained from the hydrodynamic equations, we examine how macroscopic flow and the accompanying thermodynamic reorganization together determine the preferred state.

\section{Local Mass, Momentum, Energy Conservation Laws and System Geometry}

The framework presented in this study is grounded in the fundamental laws of fluid motion. The fluid behavior is governed by the conservation of mass, momentum, and energy.

We adopt a formulation based on the assumption of local thermodynamic equilibrium. Under this assumption, the evolution of the system is described by the partial differential equations of irreversible thermodynamics \citep{deGroot2013}. These equations capture both mechanical and thermal processes in a consistent manner. To close the system, we employ the equations of state for an ideal (perfect) gas

\begin{subequations}
\label{eq:fullNSDim}
\begin{gather}
  \frac{\partial \rho}{\partial t} + \nabla \cdot \left( \mathbf{v} \rho \right)  =  0 , \label{eq:massCon}
 \\
  \rho \left[ \frac{\partial \mathbf{v} }{\partial t} +  \left( \mathbf{v} \cdot \nabla \right) \mathbf{v}   \right] 
  = - \nabla \cdot  \left( \mathbf{I} p + \bm{\Pi} \right) + \rho \mathbf{g},  \label{eq:momCon}
 \\
   \frac{\partial \left( \rho u \right) }{\partial t} =
    - \nabla \cdot \left( \rho u \mathbf{v} \right) 
    -  \left( \mathbf{I} p + \bm{\Pi} \right) : \nabla \mathbf{v} + k \nabla^2 T, \label{eq:enCon}
 \\
   \bm{\Pi} =  \frac{2}{3} \mu \left( \nabla \cdot \mathbf{v} \right) \mathbf{I} - \mu \left( \nabla \mathbf{v} + \left(\nabla \mathbf{v}\right)^{\textrm{T}} \right),
 \\
  p = \rho \frac{R T}{M}, \qquad u = \frac{3}{2} \frac{R T}{M}, \label{eq:fundRel}
 \\ 
   \frac{\partial (\rho v^2/2)}{\partial t} 
   = 
   - \nabla \cdot \left( \mathbf{v} (\rho v^2/2) \right)
   - \mathbf{v} \cdot \left( \nabla \cdot \left( \mathbf{I} p + \bm{\Pi} \right) \right)
   + \rho \mathbf{v} \cdot \mathbf{g}.
 \label{eq:kinCon}
\end{gather}
\end{subequations}
Here, $t$ is time, $\rho$ is the gas mass density, $\mathbf{v}$ is the gas velocity, $p$ is the thermodynamic pressure, $\mathbf{I}$ denotes the unit tensor, $\bm{\Pi}$ is the dynamic part of the stress tensor with viscosity $\mu$ and zero bulk viscosity, $\mathbf{g} = - g \hat{\mathbf{e}}_z$ is the gravitational acceleration vector in the direction $\hat{\mathbf{e}}_z$, $u$ is the internal energy density per unit of the mass, $T$ is the temperature, $k$ is the thermal conductivity coefficient, $R$ is the gas constant, and $M$ is the gas molar mass.

\begin{figure}[!hbt]
 \begin{center}
  \includegraphics[width=0.5\textwidth]{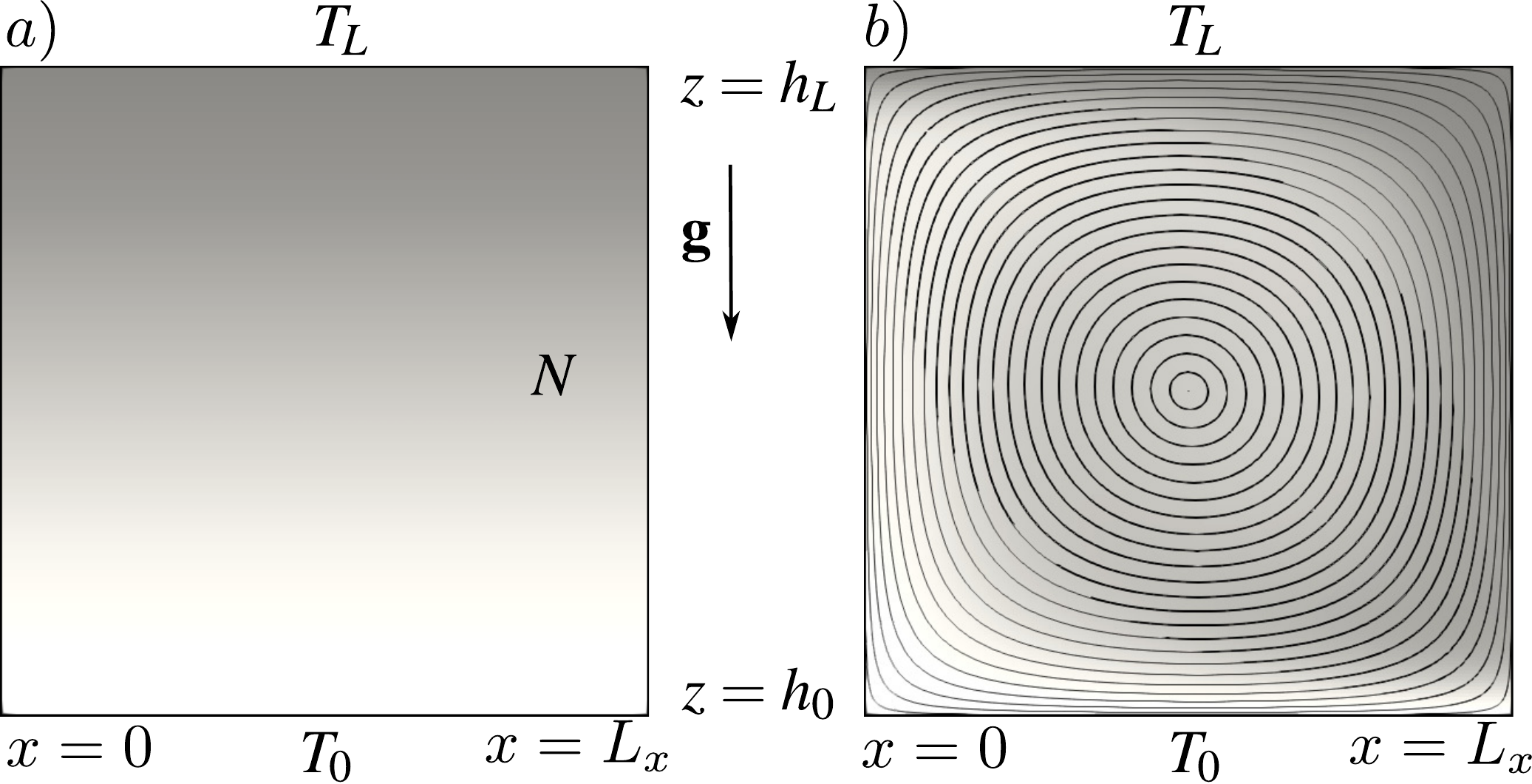}
 \end{center}
 \caption{Square domain filled with a perfect gas under gravity. The bottom wall is kept at $T_0$ and the top wall at $T_L < T_0$. The grey map corresponds to the temperature distribution. The lighter shade indicates a higher temperature. 
 Panel a) illustrates the quiescent solution that supports a heat flux without macroscopic fluid motion. 
 Panel b) illustrates a single Rayleigh-Bénard convection cell, which supports a heat flux with the occurrence of macroscopic fluid motion.
 Additionally, we present the flow lines for clockwise rotation with slip boundary conditions.}
 \label{fig:figure1}
\end{figure}

For the sake of clarity, we will restrict our discussion to a basic system geometry and focus on a 2D problem assuming translational invariance in the $y$ direction.
The considered domain is a box (figure \ref{fig:figure1}) that has base length $L_x$ spanning from $x=0$ to $x=L_x$ and height $L_z$ spanning from $z=h_0$ to $z=h_L=h_0+L_z$. 
Unless stated otherwise, we use a square box $L_z=L_x=L$. 
The domain is filled with $N$ moles of perfect gas per unit length $L_y$ in the $y$ direction.
We orient the gravitational field downwards in the $z$ direction and apply the temperature difference between the bottom $z=h_0$ and top $z=h_L$ walls.
At the side walls $x=0,\ x=L_x$, we apply the adiabatic (zero flux) boundary condition, that is $\nabla T \cdot \mathbf{n} = 0$, where $\mathbf{n}$ is a vector in the direction normal to the wall pointing outwards. 
We apply the slip ($\mathbf{v} \cdot \mathbf{n} = 0$ and zero tangetial stress) or no-slip (${\bf v}=0$) boundary conditions at all walls.
The boundary conditions for pressure are $\nabla p \cdot \mathbf{n} = 0$ at side walls and $\nabla p = \rho \mathbf{g}$ at the bottom and top walls.

\section{Quasi-static Processes}

In classical thermodynamics, the notion of quasi-static processes is applied to the transition between equilibrium states \citep{Holyst2012}. The quasi-static process is so slow that at any intermediate state between the initial and final states of the process, the system is infinitely close to the equilibrium state. In short, the quasi-static process proceeds via a sequence of equilibrium states. The same notion is needed in non-equilibrium thermodynamics. Only by studying such processes could we find the proper thermodynamic relations in the out-of-equilibrium processes. 
If the process is not quasi-static, the work done by the system and the heat transferred to and from the system differ from those in the quasi-static process, as we will see in the example below. 

The differences between the initial and final states are conveniently described by the changes of state functions, for example, the internal energy $\Delta E_T$.
The process is accompanied by the transfer of energy to the system via heat $Q$ and work $W$ from its surroundings. 
Importantly, the exact way that the change happens determines its magnitude, which satisfies $\Delta E_T = Q + W$.

Consider the classic example of a piston and cylinder filled with an ideal gas, in contact with the thermostat. 
We recall two versions of a thought experiment in which the piston is moved outwards a finite distance $x$ into the void, with the outside pressure 0. 
In the first version of the experiment, the expansion is rapid (above the velocity of sound) and uninterrupted.
Therefore, no work is being performed by the gas on its surroundings, and $W=0$.
Because the system is in contact with the thermostat, there is no change in thermal energy and $\Delta  E_T=0$ from the beginning to the end of the process.
Thus, no heat was exchanged with the surroundings and $Q=0$.

In the second version of the experiment, the gas pushes the piston quasi-statically, a process supported by the external force $F_{ext}$ that balances the gas pressure. 
During the quasi-static expansion, gas inside the cylinder has a well-defined pressure $p$. In each step, therefore, the piston performs work over the external device supporting conditions for a quasi-static process $\ddbarr W_{ext}=F_{ext} \dd x=- p \dd V$ ($\ddbarr$ denotes inexact differential).
Here, like in the first version, thermal energy is constant. 
This results in heat exchange with the environment that balances the work performed by the gas over the whole process: $Q = - W_{ext}$. 

The second law of thermodynamics $\Delta E_T - Q = W_{ext} \leq 0$ states that a gas will expand if allowed to do so. 
The second version of the thought experiment shows that the work of an external device is necessary to move the system along a trajectory consisting of stationary states.
Intuitively, a quasi-static process must be infinitely slow and at zero acceleration. Therefore, the external device used in the quasi-static process must balance the internal forces, and thus it must be equal to the change of momentum in time induced by these forces in the absence of an external device.

\section{Preliminary Example: Quiescent Gas Column in a Gravitational Field Changing Quasi-statically}
\label{sec:preliminary}

We formalize the concept of the quasi-static process by stating that during such a process, the momentum equation can be regarded as stationary $\frac{\partial \mathbf{v}}{\partial t} = 0$.
That was certainly true in the version of the thought experiment with the slow expansion of the gas, which is comprehensible within the equilibrium thermodynamics framework. 
Moreover, from studies of non-equilibrium processes, we know that pressure is the fastest variable to reach its stationary state, while temperature is the slowest, as shown in atomistic simulations \citep{holyst2008heat} and in irreversible thermodynamics \citep{babin2005evaporation}.

As a preliminary example, we analyze the consequences of the above idea in the case of a column filled with quiescent gas ($\mathbf{v}=0$, figure \ref{fig:figure1} a) subject to a very slowly-varying gravitational field 
\begin{equation} \label{eq:gForm}
 \mathbf{g}(t) = - \hat{\mathbf{e}}_z g_0 \left( 1 + \dot{\gamma} t \right)
\end{equation}
with constants $g_0$ and $\dot{\gamma}$. 
Here $ \dot{\gamma} \ll 1/\tau$, where $\tau$ is the momentum equation relaxation time scale, 
which means that the momentum equation (\ref{eq:fullNSDim}) can be treated as stationary
\begin{equation} \label{eq:statMomEq}
 \nabla p
  = 
 \rho \mathbf{g}.
\end{equation}
When the gravitational field changes its magnitude, both the pressure and the density fields adjust.
Therefore, there must exist an infinitesimally small velocity $\bm{\varv}$ satisfying the continuity equation
\begin{equation}
 \frac{\partial \rho}{\partial t} = -\nabla \cdot \left( \rho \bm{\varv} \right)\neq 0.
\end{equation}
We call $\bm{\varv}$ a quasi-static velocity, as in essence, it informs about the displacement of mass density during a quasi-static process,
which moves the system over the trajectory consisting of the stationary states.
In general, it is underdetermined but sufficient for calculating thermodynamically relevant functions of state.

In some cases, like the one at hand, the quasi-static velocity can be well approximated by the real velocity of fluid in a sufficiently slow process, which means that the thermal energy conservation law holds for $\bm{\varv}$ in unchanged form.
As a result, we can write the equation (\ref{eq:enCon}) in such a way that the stress tensor comes from the stationary momentum equation, and other velocities are quasi-static
\begin{equation} \label{eq:kinPot}
 \frac{\partial (\rho u) }{\partial t}
 = 
   - \nabla \cdot  \left( \bm{\varv} \rho u \right)
   -   p \nabla \cdot \bm{\varv}
   + k \nabla^2 T.
\end{equation}
We further integrate the thermal energy conservation law in space in the spirit of \citep{Makuch2024} and obtain
\begin{align} \label{eq:intAllCons}
    \int \dd V \frac{\partial (\rho u) }{\partial t} 
    =
 - \int \dd V \nabla \cdot \left( \bm{\varv} \rho u  \right)
 -  \int \dd V p \nabla \cdot \bm{\varv}
 + \int \dd V k \nabla^2 T.
\end{align}
On the left-hand side, we have the partial temporal derivative of the thermal energy
\begin{equation}
   \int \dd V \frac{\partial (\rho u) }{\partial t}  = \frac{\partial E_T}{\partial t}.
\end{equation}
The first term on the right-hand side is equal to 0 due to Gauss's theorem and the impermeability of the walls.
The second term can be rewritten using integration by parts as
\begin{equation}
 - \int \dd V p \nabla \cdot \bm{\varv}
 =
 - \int \dd V  \nabla \cdot \left( p \bm{\varv} \right)
 + \int \dd V \bm{\varv} \cdot \nabla p 
 =
  \int \dd V \bm{\varv} \cdot \rho \mathbf{g},
\end{equation}
where we used Gauss's theorem and impermeability of walls again, and the stationary momentum equation (\ref{eq:statMomEq}).
We recovered the power that gas particles require to move ($\bm{\varv} = \varv \hat{\mathbf{e}}_z$) in the gravitational field and therefore perform work against the gravitational force (note the negative sign)
\begin{equation}
   \dot{W}_{g,i}
 =
 - \int \dd V \rho \bm{\varv} \cdot \mathbf{g}
 =
   \int \dd V \varv \rho g_0 \left(1 + \dot{\gamma} t \right).
\end{equation}
The index $i$ denotes the work performed within the system.
Additionally, $\dot{W}_{g,i}$ can be expressed by means of a certain function of state: the potential energy of the gravitational field.
The following reasoning
\begin{align} \label{eq:eqWWEp}
  \int \dd V \varv \rho g_0 \left(1 + \dot{\gamma} t \right)
 & =
  \int \dd V \frac{\partial \left( z \varv \rho g_0 \left(1 + \dot{\gamma} t \right) \right) }{\partial z}
 -
 \int \dd V  g_0 \left(1 + \dot{\gamma} t \right) z \frac{ \partial \varv \rho }{\partial z}
 \nonumber \\
 & =  \int \dd V  g_0 \left(1 + \dot{\gamma} t \right) z \frac{ \partial \rho }{\partial t} 
 \nonumber \\
 & =  \int \dd V \frac{\partial \left( g_0 \left(1 + \dot{\gamma} t \right) z \rho \right)}{\partial t}
 - 
 \int \dd V z \rho \frac{ \partial \left( g_0 \left(1 + \dot{\gamma} t \right) \right)}{\partial t}
\end{align}
yields in the first term the rate of change of potential energy
\begin{equation}
 \int \dd V \frac{\partial \left( g_0 \left(1 + \dot{\gamma} t \right) z \rho \right) }{\partial t}
 =
 \frac{\partial E_p}{\partial t}
\end{equation}
and the second one corresponds to the negative power necessary to change the external source of gravity
\begin{equation}
 \int \dd V z \rho \frac{\partial \left( g_0 \left(1 + \dot{\gamma} t \right) \right)}{\partial t} = \dot{W}_{g,e}.
\end{equation}
In summary, the relation (\ref{eq:eqWWEp}) can be rewritten as
\begin{equation}
 \frac{\partial E_p}{\partial t} = \dot{W}_{g,i} + \dot{W}_{g,e}.
\end{equation}
Note that the determination of the quasi-static velocity is unnecessary to calculate any of the above terms, as all can be expressed through the use of density profile evolution.

The last term in equation (\ref{eq:intAllCons}) is equal to the heat flow rate through the boundaries of the system, as follows from Gauss's theorem.
\begin{equation}
 \int \dd V k \nabla^2 T =  \int \dd S k \mathbf{n} \cdot \nabla T = \dot{Q}.
\end{equation}
The above derivations, during the small time interval $\dd t$, lead to energy conservation in the following form
\begin{equation}
 \underbrace{\dd E_T }_{\dd t \int \dd V \frac{\partial (\rho u)}{\partial t} }
 = 
 \underbrace{-W_{g,i}}_{\dd t  \int \dd V \rho \mathbf{v} \cdot \mathbf{g}} 
 + 
 \underbrace{Q}_{\dd t \int \dd V k \partial_z^2 T },
\end{equation}
or equivalently expressed with the help of potential energy and external work
\begin{equation} \label{eq:consSmallWithNoEk}
 \dd E_T + \dd E_p = Q + W_{g,e}.
\end{equation}
This equation has the form known from classical thermodynamics, where the change in total energy equals the net heat added to the system and the external work done on the system.
A numerical illustration of the results described in this section can be found in Appendix \ref{sec:gravityNumerical}. 
 
\subsection{Concept of Renormalized Mass}

The decomposition of potential energy change is in line with the concept of renormalized mass $M^{*}$ \citep{Holyst2023gravity}. 
The renormalized mass $M^*$ in non-equilibrium thermodynamics is one of the parameters of state appearing in the potential energy of the system in the gravitational field:
\begin{equation}
 E_p = \frac{g L M^{*}}{2},
\end{equation}
where $L$ is the size of the system in the direction of the gravitational field and $M^{*}$ is such that, if $M$ is the total mass of the system,  $\frac{L M^{*}}{2 M}$ is the position of the center of the mass with respect to system's lowest point. 
Using this representation, the change in potential energy is written as
\begin{equation}
 \dd E_p 
 = 
  \dd \left( \frac{g L M^*}{2} \right) 
 = 
  \underbrace{\frac{g L}{2} \dd M^{*}}_{\dd t \dot{W}_{g,i}} 
  + 
  \underbrace{\frac{M^{*} L}{2} \dd g}_{\dd t \dot{W}_{g,e}}.
\end{equation}
The equivalence to expressions indicated by underbraces holds, as we observe (\ref{eq:eqWWEp}) that
\begin{equation}
 \dot{W}_{g,i} 
 = 
  \int \dd V v \rho g_0 \left(1 + \dot{\gamma} t \right)
 =
  \int \dd V  g_0 \left(1 + \dot{\gamma} t \right) z \frac{\partial \rho}{\partial t}
 =
  \frac{g L}{2} \frac{\partial M^{*}}{\partial t},
\end{equation}
and that
\begin{equation}
 \dot{W}_{g,e} 
 = \int \dd V z \rho \frac{\partial \left( g_0 \left(1 + \dot{\gamma} t \right) \right) }{\partial t}
 =
  \frac{M^{*} L}{2} \frac{\partial g}{\partial t}.
\end{equation}
Finally, the net heat $Q$ in the process of changing the gravitational field $g$ can be written in two equivalent forms:
\begin{equation}
 Q
 = T^*\dd S^*+
  \frac{gL}{2} \dd M^*= \dd E_T + \dd E_p - {W}_{g,e}.
\end{equation}

A complete set of solutions for stationary distributions of pressure and density in arbitrary conditions can be found in \citep{Holyst2023gravity}.

\section{Energy changes in the case of fluid motion and direction of spontaneous processes}

In the case of a quiescent column in a stationary state, we had no macroscopic velocity.
Now, for the first time, we will consider an example with a finite velocity $\mathbf{v}$ present in the equations. 

We showed that global non-equilibrium thermodynamics for a quiescent column is equivalent to irreversible thermodynamics when changes in the system occur sufficiently slowly that the momentum equation can be regarded as stationary. 
We extend this idea to the case that includes macroscopic motion, but first, we discuss a conceptual difficulty.
The stationarity of the momentum equation is inconsistent with the evolution of kinetic energy. 
If the process is quasi-static, i.e., at zero acceleration, how can we change the velocity?
If there is no change in velocity profile ($\frac{\partial \mathbf{v}}{\partial t}=0$), how can there be a change in kinetic energy?
Although at the level of the equations of motion, there is such a contradiction, the change in energy can be computed directly from the Navier-Stokes equations. First, we have to add
an external force, $\mathbf{F}_{ext}$, in the momentum equation
\begin{equation}
   \rho \frac{\partial \mathbf{v}}{\partial t} = 
  - \nabla \cdot \left( \mathbf{I} p + \bm{\Pi} \right)
  - \rho  \left( \mathbf{v} \cdot \nabla \right) \mathbf{v} 
  - \hat{\mathbf{e}}_z \rho g
  + \mathbf{F}_{ext}.
\end{equation}
Next, we integrate all dynamical equations to compute the change in energy, as in the preliminary example, where we analyze the quiescent gas column in a slowly varying external gravitational field (\ref{eq:gForm}).
Formal integration of energy conservation laws (\ref{eq:enCon},\ref{eq:kinCon}) over volume and time \citep{Makuch2024} leads to
\begin{equation} \label{eq:qsFree}
 \dd E_T + \dd E_k = - W_{g,i} + W_{ext} + Q,
\end{equation}
where $W_{ext}$ is the infinitesimal  work performed by the external force $\mathbf{F}_{ext}$ on the time dependent trajectory between two stationary states. 
The same result, but written in a form including an explicit change of the total energy is
\begin{equation}
\label{eq:heatForm1}
 \dd E_T 
 +
 \dd E_p
 +
 \dd E_k
 = 
 W_{g,e}
 +
 W_{ext}
 +
 Q.
\end{equation}
Finally, we have to specify $W_{ext}$ in the quasi-static process along the stationary states. 
\subsection{Determination of the external work, $W_{ext}$ moving the system along the path of stationary states}

$\mathbf{F}_{ext}$ must balance all internal forces to ensure the stationary form of the momentum equation. Usually unbalanced forces lead to a change of momentum with time. Thus such an external force is given by the momentum change in time:
\begin{equation}
 \mathbf{F}_{ext} = -\frac{\partial(\rho \mathbf{v})}{\partial t}
\end{equation}
This is a requirement for the quasi-static process going along the trajectory of stationary states at vanishing acceleration. 
The power needed to perform the work by the external force, $\mathbf{F}_{ext}$, in the quasi-static process along a stationary trajectory is as follows
\begin{align}
  \dot{W}_{ext}
  &=
  \int \dd V \mathbf{v} \cdot \mathbf{F}_{ext} 
  =
  -\int \dd V \mathbf{v} \cdot\frac{\partial (\rho\mathbf{v})}{\partial t}
  \nonumber \\
  &=
  -\int \dd V \frac{\partial (\rho v^2/2)}{\partial t}
  -
  \int \dd V \mathbf{v} \cdot \rho \left( \mathbf{v} \cdot \nabla \right) \mathbf{v} ,
\end{align}
where we have used Gauss's theorem to set $\int \dd V \nabla \cdot \left(\mathbf{v} \rho v^2 / 2 \right)=0$.

The analysis of the inertial contribution to $W_{ext}$ leads us to the Helmholtz-Hodge decomposition of the inertial term
\begin{align} 
 \int \dd V  \mathbf{v} \cdot  \rho \left( \mathbf{v} \cdot \nabla \right) \mathbf{v} 
 = 
 \int \dd V  \rho \mathbf{v} \cdot \left(  \nabla \psi + \nabla \times \mathbf{B} \right).
\end{align}
The inertial force term becomes a sum of a gradient $\psi$ and vector potential $\mathbf{B}$, both performing work due to the velocity $\mathbf{v}$. 
In Appendix \ref{sec:surWork} we show that the $\mathbf{B}$ part corresponds to the work performed at the surface since no external force moves the walls.
Next, we treat the potential term like the gravitational one
\begin{align}
 \int \dd V  \rho \mathbf{v} \cdot \left( \nabla \psi \right)
  & =
   \int \dd S  \psi \mathbf{v} \rho  \cdot \mathbf{n} 
 - \int \dd V  \psi  \nabla \cdot \left(  \rho \mathbf{v}
 \right)
 \nonumber \\
  & 
  =
  \int \dd V  \psi  \frac{\partial  \rho }{\partial t}
  =
  \int \dd V  \frac{\partial \left( \psi \rho \right)}{\partial t} - \int \dd V    \rho \frac{\partial  \psi}{\partial t}
\end{align}
and obtain a decomposition
\begin{equation} \label{eq:Psidec}
  \dot{W}_{\psi,i} + \dot{W}_{\psi,a} =  \frac{\partial \Psi}{\partial t}.
\end{equation}
It has the function $\Psi$ instead of the gravitational potential energy, 
and auxiliary work $\dot{W}_{\psi,a}$ instead of the work by the external gravity source
\begin{equation}
    \dot{W}_{\psi,i} = \int \dd V  \psi  \frac{\partial  \rho}{\partial t} = \int \dd V  \rho \mathbf{v} \cdot \left( \nabla \psi \right), \quad
    \dot{W}_{\psi,a} = \int \dd V  \rho \frac{\partial  \psi}{\partial t}, \quad 
    \frac{\partial \Psi}{\partial t} = \int \dd V  \frac{\partial \left( \psi \rho \right)}{\partial t}.
\end{equation} 
The field $\psi$ needs to be calculated to get the values of the proposed expressions.
If $\psi$ is to be interpreted as the potential field, it leads to the force $ -\nabla \psi = - \left( \mathbf{v} \cdot \nabla \right) \mathbf{v} + \nabla \times \mathbf{B}$.
Therefore, $\psi({\bf r})$ satisfy the equation:
\begin{equation}
 \nabla^2 \psi = - \nabla \cdot \left( - \left( \mathbf{v} \cdot \nabla \right) \mathbf{v} \right) = \nabla \cdot \left( \left( \mathbf{v} \cdot \nabla \right) \mathbf{v} \right) 
\end{equation}
with a boundary condition to satisfy the surface normal gradients
\begin{equation}
 \left. \nabla \psi \cdot \mathbf{n} \right|_{A} =  \left. \left( \left( \mathbf{v} \cdot \nabla \right) \mathbf{v} \right) \cdot \mathbf{n} \right|_{A},
\end{equation}
which leaves the choice of the reference value $\psi_0$ free, also called a gauge. For each stationary state, $\psi_0$ can be different.

Now we will analyze the dependence of the $W_{\psi,i}$ and $W_{\psi,a}$ terms on the gauge field $\psi_0$.
By definition, we have
\begin{align}
 \dot{W}_{\psi,a} 
 & = 
 \int \dd V  \rho \frac{\partial  \psi }{\partial t}
 = 
 \int \dd V  \rho \frac{ \partial  \left( \psi' + \psi_0  \right)}{\partial t}
 \nonumber \\
 & 
 =
 \int \dd V  \rho \frac{\partial  \psi'}{\partial t} +  \frac{\partial \left(M \psi_0\right)}{\partial t}
 =
 \dot{W}_{\psi',a} + \frac{\partial \Psi_0}{\partial t},
\end{align}
where $M$ is the total mass of the system, and $\psi'$ is the new potential shifted by the constant $\psi_0$.
Although $\psi_0$ is constant in space, it depends on the stationary state. 
Consequently, over a short time interval $\dd t$, we have
\begin{equation}
 W_{\psi,i} = \dd \Psi - W_{\psi,a} = \dd \Psi - \dd \Psi_0 - W_{\psi',a} 
 = \dd \Psi' - W_{\psi',a} = W_{\psi',i}.
\end{equation}
This result is expected: the actual infinitesimal work done by gas compression due to inertial forces does not depend on the choice of gauge. 
However, what we can see is that the gauge field can make
$W_{\psi,a}$ arbitrarily large or small. 
We choose the gauge for which $W_{\psi,a}=0$ during the transition between stationary states:
\begin{equation}
 W_{\psi,i} =  \dd \Psi - W_{\psi,a} = \dd \Psi
\end{equation}
and we arrive at the relation between the potential $\Psi$, which is a function of state, and the external work $W_{ext}$
\begin{equation}
W_{ext} = -\dd E_k-\dd \Psi.
\end{equation}
The new function of state $\Psi$ accounts for work performed on compression of the gas by inertial forces, which accompany formation of the convective state with macroscopic kinetic energy $E_k$.

The proposed gauge defines a unique quasi-static trajectory of the process, and the work done equals the change in the function of state, $-E_k-\Psi$.
The gauge can be defined at the onset of convection, where all the velocities are small. The compression of the gas by inertial forces should approach zero as $v^2/2$, i.e., as kinetic energy.
Based on the nearly marginal convection (Appendix \ref{sec:NMC}) we postulate the following gauge: $\psi=0$ at the boundary for no-slip boundary conditions and $\langle \psi \rangle =0$ at the boundary for slip boundary conditions. 

We are now in a position to calculate the external work needed to move fluid in  the Rayleigh-B\'enard cell from the initial to final stationary state: 
\begin{equation}
\label{eq:statefunction}
 \dd E_T 
 +
 \dd E_p
 +
 \dd E_k
 - 
 W_{g,e}
 -Q
=
 W_{ext}
=
 -\dd E_k - \dd \Psi=\dd\mathcal{F}.
\end{equation}
The function of state $\mathcal{F}=-E_k-\Psi$ can be calculated for any state of the system. This function allows comparison of the stability of various states in the system under the same boundary conditions. The direction of the  spontaneous processes of moving from one state to the other is set by the inequality 
\begin{equation} \label{eq:scndLaw}
\Delta\mathcal{F} \le 0. 
\end{equation}
When this function of the state decreases, the system spontaneously moves from the initial stationary state to the final one. 
The Stationary Thermodynamics provides the general form of the second law of thermodynamics (\ref{eq:scndLaw}) as a consequence of the first law.
Taking all energy differentials, subtracting the heat, and subtracting the work of external machines necessary to make a change in the system, we see that for spontaneous processes this difference is less than zero.
To summarize, in the RB convection, $\mathcal{F}=-E_k-\Psi$ is minimized, thereby setting the direction of spontaneous processes in the system.  This is illustrated below in a series of numerical experiments. 

\section{Numerical simulations}
\label{sec:allNUMS}

We show in several examples that the newly discovered potential $\mathcal{F}=-E_k-\Psi$, which is a function of state, dictates the direction of spontaneous processes in RB convection for an ideal gas (sections \ref{sec:OCnumerical}, \ref{sec:CEnumerical}, \ref{sec:CGnumerical}). Indeed, $\mathcal{F}$ decreases in spontaneous processes.

We use a custom OpenFOAM \citep{Weller1998,Jasak2007} solver, which can be downloaded from the GitHub repository \citep{githubPageSolver}, along with test cases preconfigured to facilitate reproduction of the numerical results.

In most cases, we use a square 2D simulation domain that has boundaries of length $L_x=L_z=10$ cm.
The top wall is always kept at $T_L=T_{\textrm{eq}}=293$ K. 
The bottom wall ($T_0$) is warmer, and the difference is $\Delta T = T_0 - T_L$.
The working gas is helium and is present in an amount such that at $\Delta T=0$, the inside pressure is the normal value $101325$ Pa.
The remaining boundary conditions are described in Figure \ref{fig:figure1}.

\subsection{Onset of convection}
\label{sec:OCnumerical}

\begin{figure}[!hbt]
 \begin{center}
  \includegraphics[width=\textwidth]{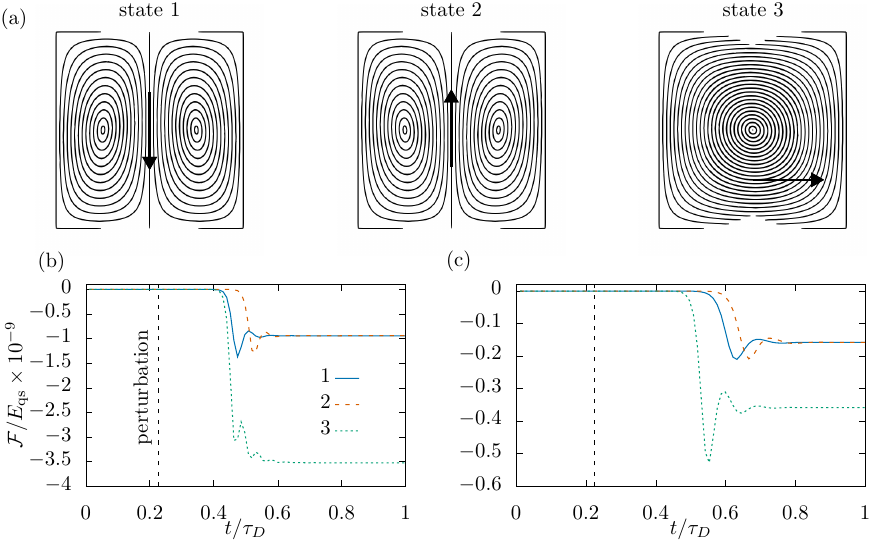}
 \end{center}
 \caption{Convection patterns appearing for $\textrm{Ra}\approx 1.6 \times 10^4$.
   a) Flow lines. The flow directions are indicated by the large arrows. The resulting convection patterns are directly correlated with the specific form of the initial perturbation. 
   b) The evolution of $\mathcal{F}$ in time for all three states shown in (a) for slip boundary conditions: the lowest value of $\mathcal{F}$ is obtained for a single-roll state 3.
   c)  The evolution of $\mathcal{F}$ in time for all three states shown in (a) for no-slip (stick) boundary conditions: the lowest value of $\mathcal{F}$ is obtained for the single-roll state 3.
   Time is normalised by the thermal relaxation scale $\tau_D = L^2 \rho_0 c_v/k \approx 88$~s and $\mathcal{F}$ is normalized by the thermal energy of the quiscent solution $E_{\textrm{qs}}$
  }
 \label{fig:cells}
\end{figure}

The first case that we study is the onset of convection from the quiescent state at $\Delta T = 5$ K, which gives Rayleigh number $\textrm{Ra} \approx 1.6 \times 10^4 > \textrm{Ra}_c$ (Appendix \ref{sec:RN}). 
The critical Rayleigh number $\textrm{Ra}_c$ for a single convective roll is $\textrm{Ra}_c\approx 779$ for slip walls and $\textrm{Ra}_c\approx 2585$ for no-slip (stick) walls.
The bifurcation point for the double convective roll is $\textrm{Ra}_c \approx 3044$ for slip walls and $\textrm{Ra}_c\approx 6742$ for stick walls (figure \ref{fig:cells}a).
The proposed conditions also satisfy the reported stability for the two-roll solution with stick walls \citep{venturi2010}.
As a result, both the single and double-roll states shown in Fig(\ref{fig:cells}a) are stable. 
In Fig(\ref{fig:cells}b), we show $\mathcal{F}$ as a function of time after perturbing the quiescent state. 
The lowest value of $\mathcal{F}$ is for the single-roll, which is the most stable solution (both for slip and no-slip boundary conditions).

States 1 and 2 correspond to double convection cells, while case 3 corresponds to a single one.
There is also the quiescent state for an extended period of time (with the numerical scheme used), provided no perturbation is introduced.
The system was initiated in a quiescent state.
Next, we solve the full 2D set of equations unperturbed until we introduce a small perturbation, which induces the emergence of convection cells.
The perturbation was a point-velocity source applied over a single time step. The perturbation was different for the final stationary single- and double-roll states.
We investigate the change in $\mathcal{F}$, in time (figure \ref{fig:cells}b). The smallest value of $\mathcal{F}$ was reached for state 3, a single-roll, which is the most stable solution for a quadratic box.

There are other suggestions that state 3 is the most stable. For example, the bifurcation to state 3 occurs at lower Ra numbers than the bifurcation to the double-roll states 1 and 2. Its stability can also be argued based on the ordering of magnitudes of eigenvalues of modes during the linear stability analysis \citep{boulle2022bifurcation}. 
However, these arguments lack the thermodynamic interpretation, which we have presented here.

\subsection{Changes of external parameters}
\label{sec:CEnumerical}

\begin{figure}[!hbt]
 \begin{center}
  \includegraphics[width=\textwidth]{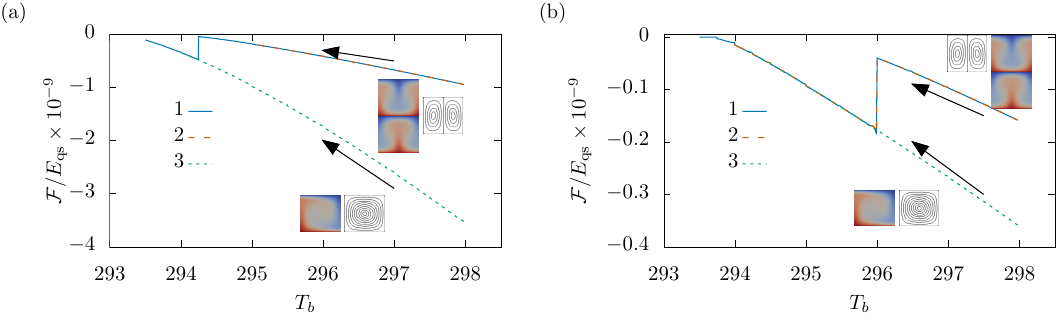}
 \end{center}
 \caption{$\mathcal{F}$ as a function of $T_b$ during slow changes of $T_b$.
 a) slip boundary conditions on all walls
 b) no-slip (stick) boundary conditions on all walls.
 $\mathcal{F}$ is normalized by thermal energy of quiscent state $E_{\textrm{qs}}$ at $\Delta T = 5$~K.}
 \label{fig:changeT}
\end{figure}

\begin{figure}[!hbt]
 \begin{center}
  \includegraphics[width=\textwidth]{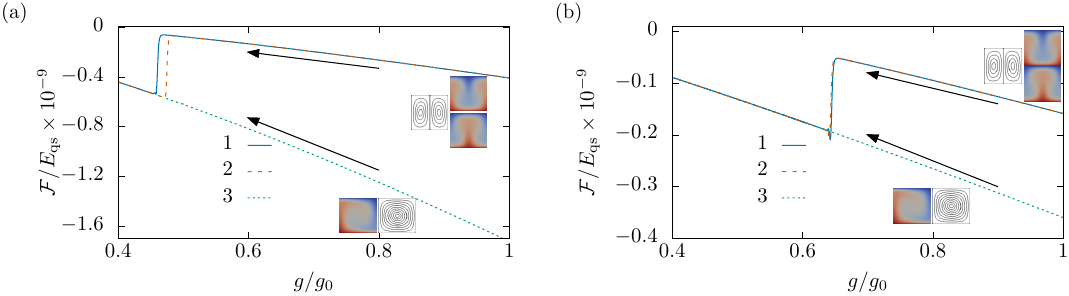}
 \end{center}
 \caption{$\mathcal{F}$ as a function of $g$ during the slow changes of $g$ normalized by initial $g_0$.
   a) Slip boundary conditions on all walls. Here a lower $\Delta T = 3$ K, which gives $\textrm{Ra} \approx 9.6 \times 10^3$, supporting all three states of convection cell.
   b) No-slip (stick) boundary conditions on all walls.
   $\mathcal{F}$ is normalized by thermal energy of quiscent state $E_{\textrm{qs}}$ at $\Delta T = 3$~K.
  }
 \label{fig:changeG}
\end{figure}

We induce the transition between states 1,2 and state 3 by changing the temperature of the bottom reservoir (Figure \ref{fig:changeT}) and by changes in the magnitude of the gravitational field (Figure \ref{fig:changeG}).
We initiated the system in stationary states 1, 2, and 3.
Next, we performed slow, continuous, time-resolved sweeps that varied the magnitude of the respective external parameter.
We intended to remain close to the stationary state, but did not wait for the system to relax completely. A comparison of the slow time-resolved sweep to the stationary solution showed differences in $\mathcal{F}$ value below 0.1\%.

In all cases, we observed a spontaneous transition where a convection cell in state 1 or 2 changed into a convection cell in state 3.
We never observed a spontaneous transition in the other direction, indicating that state 3 is indeed the most stable.
Moreover, every transition was always associated with a decrease of $\mathcal{F}=-E_k-\Psi$.

For the stick BC, the system changed from the double-roll state to a single-roll before reaching $\textrm{Ra}_c$, which is caused by the instability region just above the $\textrm{Ra}_c$ value, for which the double-roll convection pattern appears \citep{venturi2010}. For slip BCs, the transition varied with the forcing. Decreasing $T_b$ led to almost a stop of the flow, while decreasing $g$ led to a transition that occurred in an apparently smoother way, with more pronounced distinction between the two directions of velocity rotation. 

\subsection{Changes of geometry}
\label{sec:CGnumerical}

\begin{figure}[!hbt]
 \begin{center}
  \includegraphics[width=\textwidth]{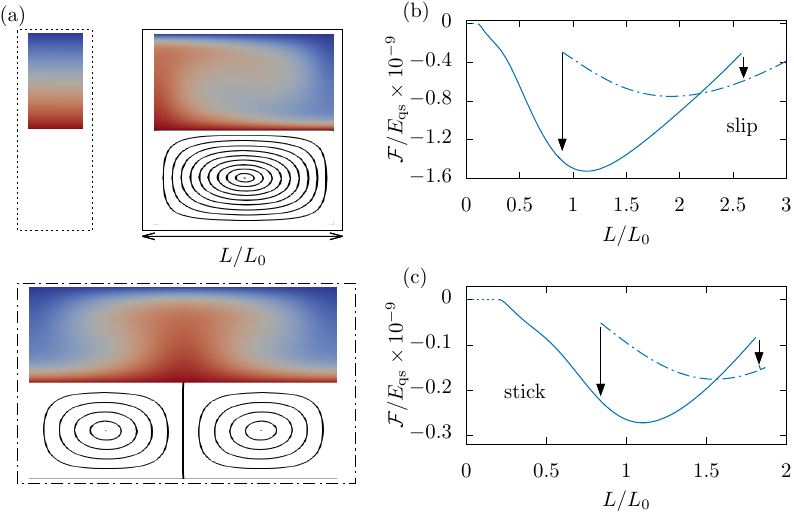}
 \end{center}
 \caption{Spontaneous transitions during changes in the shape of the box $L_x$. 
   a) Typical features of three competing states. The temperature profile as a heat map, ranging from cold (blue) to hot (red), with flow lines. 
   b) The value of $\mathcal{F}$ for different sizes of the system and all three states. For the quiescent state $\mathcal{F}=0$. The directions of spontaneous changes of state and changes in $\mathcal{F}$ are indicated by the arrows. The solid line is $\mathcal{F}$ as a function of $L_x$ for a single-roll state, while the dashed-dotted line is $\mathcal{F}$ for the double-roll state. The transition should occur at the intersection of these two lines, but due to the state's metastability, they persist beyond the crossover point. Here, we set slip boundary conditions. The spontaneous transitions are indicated by arrows. 
c) same as in (b), but for no-slip (stick) boundary conditions
   A~movie~(supplementary information 1) illustrates the complete expansion and compression cycle.
  }
 \label{fig:changeLx}
\end{figure}

Another way to induce a change of state is by altering the shape of the simulation domain (Figure \ref{fig:changeLx}). 
We initiated the simulation in state 1 for a square box,
and varied the box shape ($L_x$), keeping all other parameters fixed.
We used $\Delta T = 3$ K for slip BCs and $\Delta T = 4$ K for stick BCs.
After each small change of the box shape, we simply stretched (or squeezed) the previous solution and allowed it to relax to the stationary state.
Therefore, this is a record of stationary states, not a fully resolved dynamic sweep, as when we changed the strength of gravity. 
In Figure \ref{fig:changeLx}, we showed three competing states: quiescent state, a state with a single-roll, and a state with a double-roll. For a very small box size, the convection states are not stable. As we increase the size of the box, a single-roll appears; for larger sizes, two rolls become the most stable solution. For some box sizes, the single-roll and double-roll are stable, and the lowest value of $\mathcal{F}$ determines which is more stable.
We observed two spontaneous changes.
The first transition was continuous from the quiescent state to the single-roll state. Once the box was sufficiently elongated, we observed a spontaneous transition from the single-roll state to the double-roll state. 
Upon reaching a stable solution in the double-roll state, we began decreasing the size of the box.
Then, the second spontaneous transition occurred once the box became sufficiently small from the double-roll state to the single-roll state.
In both cases, we observed a decrease in $\mathcal{F}=-E_k-\Psi$.
We did not observe spontaneous transitions in the opposite direction with increase of $\mathcal{F}$.

\section{Conclusions}

In this work, we have developed a global non-equilibrium thermodynamic description of stationary fluid flows in which transitions between distinct stationary states are associated with a path-independent thermodynamic cost. Starting from the governing hydrodynamic equations and the corresponding global balance relations, we showed that this cost can be written as the change of a state function,
\begin{equation}
\mathcal{F}=-E_k-\Psi,
\end{equation}
where $E_k$ is the macroscopic kinetic energy and $\Psi$ is a thermodynamic state function emerging directly from the nonlinear inertial term of the Navier–Stokes equation. 
$\Psi$ is defined by
\begin{equation}
\Psi=\int_V \rho({\bf r})\psi({\bf r}) \dd V,
\end{equation}
where the integral is taken over the system volume for a given stationary state. 
The scalar field $\psi ({\bf r})$ is obtained from the Helmholtz–Hodge decomposition of the inertial term $\left(\mathbf{v}({\bf r}) \cdot \nabla \right) \mathbf{v}({\bf r})$. 
The field $\psi$ is defined up to an arbitrary constant at each instant of time during a transition between stationary states. 
With a proper choice of gauge for $\psi$, one finds that the total external work required to transfer the system from one stationary state to another (under constant boundary conditions) is given by
\begin{equation}
W_{ext} =\Delta\mathcal{F}=-\Delta E_k -\Delta\Psi,
\end{equation}
which depends only on the initial and final stationary states, and not on the specific trajectory of the process. 
Thus, $\Delta\mathcal{F}$ is the work done by the external device on the trajectory consisting of stationary states to change the macroscopic flow pattern from the initial to the final pattern. 

The resulting formulation provides a global criterion for comparing stationary flow states realized under the same external constraints. Among dynamically accessible stationary states, lower values of $\mathcal{F}$ correspond to more stable states and determine the preferred direction of spontaneous transitions. The kinetic and thermodynamic contributions act in the same direction: both the development of macroscopic kinetic energy, and the accompanying compression and redistribution of the fluid, serve to reduce $\mathcal{F}$. When $\mathcal{F}$ decreases, the stability of the state increases. The corresponding inequality, therefore, constrains transitions between stationary flows without requiring knowledge of the detailed dynamical path connecting the initial and final states.

We tested this criterion using two-dimensional Rayleigh-B\'enard convection, in which several stationary flow configurations can coexist under the same control parameters. The numerical results show that the states selected by the dynamics are consistently those with lower values of $\mathcal{F}$. This behavior is observed for both slip and no-slip boundary conditions, indicating that the thermodynamic selection criterion is not specific to a particular momentum boundary condition. Rayleigh-B\'enard convection thus provides a direct hydrodynamic realization of the general thermodynamic construction.

An important aspect of the present formulation is that the state function is derived directly from the hydrodynamic equations. In particular, $\Psi$ originates from the nonlinear inertial term responsible for coupling between the macroscopic velocity field and the accompanying redistribution of the fluid. The resulting quantity $\mathcal{F}=-E_k-\Psi$ therefore combines the macroscopic motion and the thermodynamic reorganization required to sustain that motion within a single function of state. This establishes a direct connection between the nonlinear structure of the Navier-Stokes equation and a global thermodynamic criterion for stationary-flow stability.

The criterion developed here is complementary to conventional hydrodynamic stability analysis and bifurcation theory. These latter approaches determine whether a given flow state is dynamically stable and describe how instabilities develop as control parameters are varied. The present formulation addresses a different question: {\it when several stationary states are dynamically accessible under identical external constraints, can their relative stability and preferred direction of transition be inferred from the states themselves?} The state function $\mathcal{F}$ provides such a global measure, without requiring the transition trajectory to be resolved. 

In this paper, we have discussed compressible flows. In the incompressible limit, the function of state simplifies to $\mathcal{F}=-E_k$, since at constant density there is no extra work in compression of the gas. There are already some results indicating that this works for the transitions in incompressible flows, i.e., that more stable states (at constant boundary conditions) exhibit larger kinetic energy, i.e., smaller $\mathcal{F}$ \citep{lamee2015,Bars2021}.

The present calculations have been restricted to stationary two-dimensional Rayleigh-B\'enard convection, in which the thermodynamic structure can be examined under controlled conditions. An important next step will be to determine how broadly the same construction applies to three-dimensional convection and to other stationary flows with more complex geometries and forcing. More generally, the existence of a path-independent state function for macroscopic flow reorganization suggests that stationary hydrodynamic systems may possess a global thermodynamic structure that complements the local equations governing their dynamics. The results presented here provide a concrete example of such a structure and a thermodynamic criterion for selecting stationary macroscopic flow states.

\section*{Acknowledgements}
We thank Karol Makuch for the insightful discussion.

\section*{Funding statement}
This research received no specific grant from any funding agency, commercial or not-for-profit sectors.

\section*{Data availability statement}
Data supporting the findings of this study are available for download from the GitHub repository \citep{githubPageSolver} and have a Zenodo DOI.

\section*{Declaration of interest}
The authors report no conflict of interest.

\section*{Author ORCID}
R. Ho\l yst, https://orcid.org/0000-0002-3211-4286,

P. J. \.Zuk, https://orcid.org/0000-0003-0555-5913,

K. Gi\.zy\'nski, https://orcid.org/0000-0002-3547-1174,

A. Macio\l{}ek, https://orcid.org/0000-0002-2371-4183,

P. V. E. McClintock, https://orcid.org/0000-0003-3375-045X

\section*{Author contributions}
R. Ho\l yst and P. J. \.Zuk contributed equally.

\appendix

\section{Toy model}
\label{sec:tm}

\begin{figure}[!hbt]
 \begin{center}
  \includegraphics[width=0.5\textwidth]{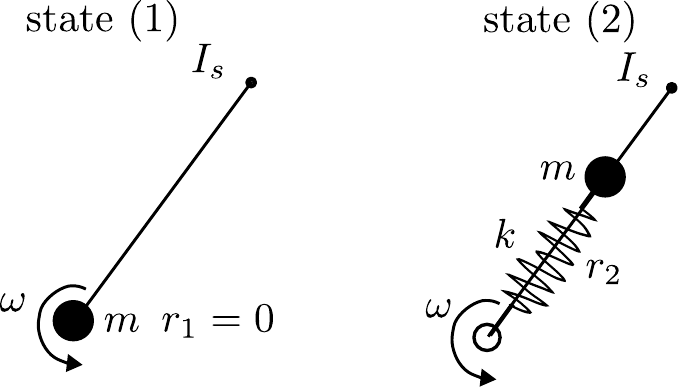}
 \end{center}
 \caption{Toy model consisting of a rotating stick and a point mass attached to a nonlinear spring. 
 The model has two stationary states: (1) an unstable state with the spring compressed, and (2) a stable state with the spring extended.}
 \label{fig:toyModel}
\end{figure}

To build intuition for the results presented in this paper, we introduce a simple toy model: 
a rotating stick with moment of inertia $I_{s}$ about its rotation axis, and a point mass $m$ attached to the center of rotation by an anharmonic spring with restoring force $F(r) = -k r^3$, where $k$ is the spring constant. 
There is no gravitational field in this model.
The point mass can slide along the stick (figure \ref{fig:toyModel}). 

We assume that the stick rotates at a constant angular velocity $\omega$. 
The mass can occupy two stationary positions along the stick $(r \ge 0)$:  
(1) at the center, $r_1 = 0$, and  
(2) at $r_2 = \sqrt{\tfrac{m \omega^2}{k}}$, determined by the balance of centrifugal and spring forces. 

Let us now evaluate the system’s energies in these two states.  
In state (1) the energy is simply the rotational kinetic energy of the stick:
\begin{equation}
E_1 = \frac{I_{s} \omega^2}{2}.
\end{equation}

In state (2), the total energy is
\begin{equation}
E_2 = \frac{I_{s} \omega^2}{2} + \frac{m r_2^2 \omega^2}{2} + \frac{k r_2^4}{4},
\end{equation}
which consists of the stick's kinetic energy, the rotating point mass's kinetic energy, and the spring's elastic energy. 
The third contribution is crucial: it represents the work required to stretch the spring. 
This work is not directly provided by the device, maintaining a constant $\omega$, since that device only balances tangential forces. 
Nevertheless, it must be supplied for the system to reach state (2). 

The key observation is therefore the following:  the motion of a complex system cannot be accounted for solely by its kinetic energy. 
In general, additional work is required—in this case, the work needed to stretch the spring, which accompanies changes in the system's kinetic energy.

\section{Rotational Term in the Helmholtz-Hodge Decomposition of Inertia: Surface Contribution}
\label{sec:surWork}

We showed in a preliminary example that during a quasi-static process, the density profile changes slowly. Consequently, on top of the stationary velocity $\mathbf{v}^{st}$ there is an additional velocity $\bm{\varv}$ in the system
\begin{equation}
 \mathbf{v} = \mathbf{v}^{st} + \bm{\varv},
\end{equation}
which is necessary to satisfy the continuity equation on the stationary states, when we move infinitesimally slowly between them:
\begin{equation}
 \frac{\partial \rho}{\partial t} = - \nabla \cdot \left( \rho^{st} (\mathbf{v}^{st} + \bm{\varv}) \right).
\end{equation}

Inside a closed system, the stationary velocity cannot perform work against inertial forces, which it creates, because
\begin{align}
 \int \dd V  \rho^{st} \mathbf{v}^{st} \cdot \left( \mathbf{v}^{st} \cdot \nabla \right) \mathbf{v}^{st} 
  & =
 \int \dd V \rho^{st} \mathbf{v}^{st} \cdot \nabla \frac{(v^{st})^2}{2}
 \nonumber \\
  & =
 \int \dd A \mathbf{n} \cdot \mathbf{v}^{st} \rho^{st} \frac{(v^{st})^2}{2}
 - 
 \int \dd V  \frac{(v^{st})^2}{2} \nabla \cdot \left( \rho^{st} \mathbf{v}^{st} \right)
 \nonumber \\
 & = 0.
\end{align}
As a result, we have 
\begin{equation}
 \int \dd V  \mathbf{v} \cdot \rho^{st} \left( \mathbf{v}^{st} \cdot \nabla \right) \mathbf{v}^{st} 
   =
 \int \dd V  \bm{\varv} \cdot  \rho^{st} \left( \mathbf{v}^{st} \cdot \nabla \right) \mathbf{v}^{st}.
\end{equation}
The inertial forces come from the stationary velocity profiles, and mass is moving with quasi-static velocity $\bm{\varv}$.
We rewrite the velocity inertial acceleration, performing its Helmholtz-Hodge decomposition
\begin{align} 
 \int \dd V  \bm{\varv} \cdot  \rho \left( \mathbf{v}^{st} \cdot \nabla \right) \mathbf{v}^{st} 
 = 
 \int \dd V  \rho \bm{\varv} \cdot \left(  \nabla \psi + \nabla \times \mathbf{B} \right).
\end{align}

Next, we calculate the power needed to perform work due to the velocity $\bm{\varv}$ and the vector potential $\mathbf{B}$.
In the 2D case at hand, we find it convenient to use the relation \citep{Bhatia2012}
\begin{equation}
 \nabla \times \mathbf{B} = \mathbb{J} \nabla \chi,
 \qquad
 \mathbb{J} = 
 \left[
 \begin{array}{cc}
  0 & 1 \\
  -1 & 0
 \end{array}
 \right]
\end{equation}
with $\chi$ being a scalar function.
Upon substitution, we find that
\begin{equation}
 \int \dd V  \rho \bm{\varv} \cdot \nabla \times \mathbf{B}
 =
 \int \dd V  \rho \bm{\varv} \cdot \mathbb{J} \cdot \nabla \chi
 =
 \int \dd V  \rho \bm{\varv}^{\perp} \cdot \nabla \chi,
\end{equation}
where $\bm{\varv}^{\perp} = \bm{\varv} \cdot \mathbb{J}$ is the $\bm{\varv}$ rotated clockwise by $\pi/2$.
Next, integrating by parts, we find that
\begin{equation}
 \int \dd V  \rho \bm{\varv}^{\perp} \cdot \nabla \chi
 =
 \int \dd A \mathbf{n} \cdot \rho \bm{\varv}^{\perp} \chi
 -
 \int \dd V  \chi  \nabla \cdot \left( \rho \bm{\varv}^{\perp} \right)
 = 
 \int \dd A \mathbf{n} \cdot \rho \bm{\varv}^{\perp} \chi,
\end{equation}
where we used the identity
\begin{equation}
 \nabla \cdot \left( \rho \bm{\varv}^{\perp} \right)
 = 
 \nabla \rho \times \bm{\varv} + \rho \nabla \times \bm{\varv} 
 = 0.
\end{equation}
It can be argued first that $\bm{\varv}$ is caused by the changes in mass distribution; therefore, it must be in the direction of the density gradient ($\nabla \rho \times \bm{\varv}=0$).
Secondly, the exact value of $\bm{\varv}$ can be determined only through the mass conservation law
$$
\partial_t \rho = - \nabla \cdot \left( \rho \bm{\varv} \right),
$$
which does not give a unique recipe for the solution, like in the 1D case. However, under the further assumption that $\bm{\varv} = \nabla \xi$, a set of acceptable solutions can be reduced to such that $\nabla \times \bm{\varv}=0$. Note that this is an intrinsic property of 1D solutions.

In order to proceed further, we need to solve for $\chi$. The equation for $\chi$ is given by
\begin{equation}
 \nabla^2 \chi = - \nabla \cdot \mathbb{J} \left( \mathbf{v}^{st} \nabla \mathbf{v}^{st} \right)
 =
 - \partial_x \left(v_x \partial_x v_z + v_z \partial_z v_z \right) + \partial_z \left( v_x \partial_x v_x + v_z \partial_z v_x \right)
\end{equation}
with boundary conditions
\begin{equation}
 \left. \nabla \chi \right|_A 
 = 
 - \left. \mathbb{J} \left( \mathbf{v}^{st} \nabla \mathbf{v}^{st} \right) \right|_A.
\end{equation}
We look for $\chi$ at the domain boundary under the assumption that it is a continuous function.
First, consider a wall, which is a part of the boundary that has a normal outward vector $\hat{\mathbf{e}}_x$ (or fixed $x$ coordinate and varying $z$ coordinate).
There $v_x=0$, but also due to the slip boundary conditions $\partial_x v_z = 0$ so that on that wall, with a fixed $x$ coordinate
\begin{equation}
 \partial_z^2 \chi 
 = 
 0
\end{equation}
The same is true for the other walls.
We would like $\chi$ to be continuous in space, 
and thus we are left with $\chi=\textrm{const.}$ as a solution.
We summarize this as
\begin{equation}
 \int \dd V  \rho \bm{\varv} \cdot \nabla \times \mathbf{B}
 = 
 \int \dd A \mathbf{n} \cdot \rho \bm{\varv}^{\perp} \chi_0
 =
 \dot{W}_{\chi}
\end{equation}
where $\chi_0$ is a constant. We postulate this constant to be 0.

\section{Nearly marginal convection for slip boundary conditions}
\label{sec:NMC}
We demonstrate that, for vanishing rotations, in a leading order of expansion near the onset of convection,  $\Psi \rightarrow 0$. In this way, we choose the proper gauge for the system. 

\subsection{Functional form of the solution}

In what follows, we use the exact form of equations that are given by \citep{Mizerski2021}, and for the sake of clarity, we write them below using the notation of the present manuscript.

To study the development of a small perturbation of an arbitrary quantity $\phi$ on top of the quiescent state, we write
\begin{equation}
 \phi = \phi_{\textrm{qs}} + \phi'  
\end{equation}
where $\phi'$ is a small perturbation around the quiescent solution $\phi_{\textrm{qs}}$.
The solution proposed by Lord Rayleigh \citep{rayleigh1916lix} for normal modes has the following (Fourier-like) form (we will take the real part)
\begin{gather}
 \phi' =  \Re \varphi (z) e^{\sigma t} e^{i \mathcal{K} x}.
\end{gather}
Due to the geometic constraint $\mathcal{K} = \frac{m \pi}{L_x}$, where $L_x = \frac{x_L-x_0}{h_L - h_0}$.
Moreover, we will be interested here in solutions with a critical value of $\sigma=0$ that obey the principle of change of stability \citep{Chandrasekhar2013}.
The considered equation for the vertical velocity amplitude is
\begin{equation}
 \left(\frac{\dd^2}{\dd z^2} - \mathcal{K}^2 \right)^3 v_z = - \frac{\mathrm{Ra}}{L_z^4}\mathcal{K} v_z.
\end{equation}
We will supplement it with the appropriate boundary conditions, which means free boundaries at the top and bottom
\begin{equation}
 v_z(z=0) = 0, \ v_z(z=L_z)=0, \ \partial^{2 n}_z v_z(z=0) = 0, \ \partial^{2 n}_z v_z(z=L_z) = 0
\end{equation}
and we obtain a solution of the following form
\begin{subequations}
\begin{gather}
 v_x = - A \frac{n \pi}{L_z} \cos \left( \frac{n \pi}{L_z} z \right) \sin\left(\frac{m \pi}{L_x} x\right),
 \\
 v_z = A \frac{m \pi}{L_x} \sin\left( n \pi z\right) \cos \left(\frac{m \pi}{L_x} x \right),
 \\
 T =  A \frac{\frac{m \pi}{L_x}}{(\frac{m \pi}{L_x})^2 + (\frac{n \pi}{L_z})^2}\sin \left( n \pi z \right) \cos\left(\frac{m \pi}{L_x} x\right),
\end{gather}
\end{subequations}
with $A$ being the proportionality constant for a given mode, which cannot be determined within the scope of linear theory.
The latter theory is, however, sufficient to determine the onset point for convection, which happens at the critical Rayleigh number
\begin{equation}
 \mathrm{Ra}_c = \frac{\left((\frac{n \pi}{L_z})^2 + (\frac{m \pi}{L_x})^2\right)^3}{(\frac{m \pi}{L_x})^2}.
\end{equation}
In the special case of $L_x=L_z=1$ (in units of L),
\begin{gather} \label{eq:stabilityT}
 \mathrm{Ra}_c = \frac{\left( n^2 \pi^2 + m^2 \pi^2 \right)^3}{ m^2 \pi^2}
 \\
 \mathrm{Ra}_c = 8 \pi^4 \approx 779.27, \quad m,n=1.
\end{gather}
The linear theory is not sufficient to study stationary states, which require a perturbation series in a small parameter
\begin{equation}
 \eta = \frac{\mathrm{Ra} - \mathrm{Ra}_c}{ \mathrm{Ra}_c }.
\end{equation}
Upon substitution, following calculations \citep{Mizerski2021} this gives 
\begin{equation}
 A = \pm \frac{2 \sqrt{2} \sqrt{ (\frac{n \pi}{L_z})^2+(\frac{m \pi}{L_x})^2}}{\frac{m \pi}{L_x}}
\end{equation}
and results in corrections beyond the hydrostatic solution
\begin{align}
 v_x & = - \eta^{1/2} \kappa \frac{A n \pi}{L_z} \cos\left(\frac{n \pi}{L_z} z\right) \sin \left( \frac{m \pi}{L_x} x \right) + \ldots
 \\
 v_z & = \eta^{1/2} \kappa \frac{A m \pi}{L_x} \sin \left(\frac{n \pi}{L_z} z \right) \cos \left(\frac{m \pi}{L_x} x \right) + \ldots
 \\
 T & = 
 \eta^{1/2} (T_0-T_L) \frac{A m \pi L_z L_x}{(m \pi L_z)^2 + (n \pi L_x)^2} \sin \left( \frac{\pi  n}{L_z} z \right) \cos \left(\frac{\pi  m}{L_x} x \right) 
 - 
 \eta (T_0-T_L) \frac{1}{\pi n} \sin \left(\frac{2 \pi  n}{L_z} z \right) + \ldots 
  \\
 p & = \eta^{1/2} \frac{\rho_{\textrm{av}} \kappa^2}{L_z^2} A \textrm{Pr}_c \frac{n}{m} L_x L_z \left(
 \left( \frac{n \pi}{L_z} \right)^2+ \left(\frac{m \pi}{L_x}\right)^2\right) \cos \left(\frac{\pi  n}{L_z} z \right) \cos \left(\frac{\pi m }{L_x} x\right)
 +
 \nonumber \\
 & \eta \frac{\rho_{\textrm{av}} \kappa^2}{L_z^2} \left( 
 \frac{\pi ^2 A^2}{4} \left(n^2 \cos \left(\frac{2 \pi  m }{L_x}x\right)+ m^2 \frac{L_z^2}{L_x^2} \cos \left(\frac{2 \pi  n}{L_z} z\right)\right)+\frac{\textrm{Pr}  \textrm{Ra}_c}{2 \pi ^2 n^2} \cos \left(\frac{2 \pi  n}{L_z} z\right)
 \right) 
 +
 \ldots,
 \end{align}
where $\kappa = \frac{k}{\rho c_p}$. We can now calculate 
\begin{equation}
 \psi 
 = 
 - \eta \kappa^2 A^2 \frac{n^2 m^2 \pi^4}{4 L_x^2 L_z^2}\left[ \left( \frac{L_x}{m \pi} \right)^2 \left(\cos \left( \frac{2 m \pi x}{L_x} \right) - 1 \right) 
 + \left( \frac{L_z}{n \pi} \right)^2 \left( \cos\left(\frac{2 n \pi z}{L_z}\right)- 1 \right) \right] 
 - 
 \frac{\int \rho_{av} v^2/2 \dd V}{\int \rho_{av} \dd V}
\end{equation}
using the proposed gauge. 

\subsection{Functional form of thermodynamic potentials}

From the previous equation we can calculate the kinetic energy
\begin{align}
 E_k^{\eta} & = \int \rho \frac{v_x^2 + v_z^2}{2} \dd V 
 = \eta \rho_{\textrm{av}} \kappa ^2 \frac{\pi^2 L_y A^2 \left(L_x^2 n^2+ L_z^2 m^2\right)}{8 L_x L_z}
 + \ldots
\end{align}
and potential $\Psi$
\begin{align}
 \Psi^{\eta} & = 
  \int \rho \psi \dd V 
 = E_k^{\eta}
 + o(\eta).
\end{align}
It has a value of $E_k^{\eta}$ at leading order in the expansion. 
The work due to compression is proportional to the kinetic energy. That is why, in the leading order, both vanish. This example allows us to determine the gauge field. The gauge is such that, at vanishing velocity, both quantities vanish as $v^2$.  
Similar calculations can be done for no-slip boundary conditions.

\section{Rayleigh number in simulations}
\label{sec:RN}
 
We represent the forcing $\Delta T$ in a dimensionless form using the Rayleigh number, comparing energy transport time scales due to convection and diffusion
\begin{align}
 \textrm{Ra} &
             = \frac{L^2 / \alpha}{\mu / ( \Delta \rho L g ) }
             = \frac{L^3 g}{\mu \alpha} \Delta \rho
             = \frac{L^3 g }{\textrm{Pr} \kappa^2}\frac{\Delta \rho}{\rho_{eq}}
             \approx \frac{L^3 g}{\mu \alpha} \rho_{\textrm{eq}}\frac{T_0 - T_L}{\log T_0 - \log T_L} \left( \frac{1}{T_L} - \frac{1}{T_0} \right)
             \nonumber \\
             &= \frac{L^3 g \textrm{Pr}}{(\mu/\rho_{\textrm{eq}})^2} \frac{T_0 - T_L}{\log T_0 - \log T_L} \left( \frac{1}{T_L} - \frac{1}{T_0} \right).
\end{align}
Here, $\textrm{Pr} = \frac{\mu c_p}{\kappa}$, $\kappa = \frac{k}{\rho c_p}$, $\alpha$ is thermal diffusivity .
The last approximation is made based on the density profile from the 1D case when gravity is negligible \citep{Holyst2023fundamental,Holyst2023gravity}.
In the considered parameter regime, the latter can be further approximated by a linear expansion of $T_0$ around $T_L$ with negligible error.

\section{Numerical simulations and the First Law in quiescent column}
\label{sec:gravityNumerical}

Important questions arise from the different nature of the quasi-static process and free evolution.
Can the presented theory be verified directly?
What is the role of numerical simulations in such verification?

We investigate the first law of stationary thermodynamics numerically and examine changes in potential and work along trajectories that approximate the quasi-static process.

For the quiescent column ($E_k = 0$), temporal evolution, which is performed sufficiently slowly, appears to be equivalent to the quasi-static process.
We illustrate this by numerical simulations of the quiescent gas column under a slow change in the gravitational field.

In essence, it is a 1D simulation where all fluxes and gas displacement happen in the $z$ direction only (figure \ref{fig:figure1}a).
The simulation domain has a length $L_z=10$ cm.
The working gas has the parameters of helium.
The amount of gas in the domain is such that, at a temperature of $293$ K, the normal pressure inside is $101325$ Pa.
To reduce the total thermal energy, we keep the top and bottom walls at the low temperature, $T_L=T_0=T_{\textrm{eq}}=29.3$ K.
In case of unequal temperatures at the boundary, even a
minute gas displacement results in large changes in thermal energy in comparison to the work of gravitational energy 
\begin{equation} \label{eq:wginum}
 W_{g,i} = \int_{t_1}^{t} \dd t \int \rho g \hat{\mathbf{e}}_z \cdot \mathbf{v} \dd V
\end{equation}
upon changing $g$,
which we aim to observe.
Note that Figure \ref{fig:figure1} describes different boundary conditions.

\begin{figure}[!hbt]
 \begin{center}
  \includegraphics[width=1.\textwidth]{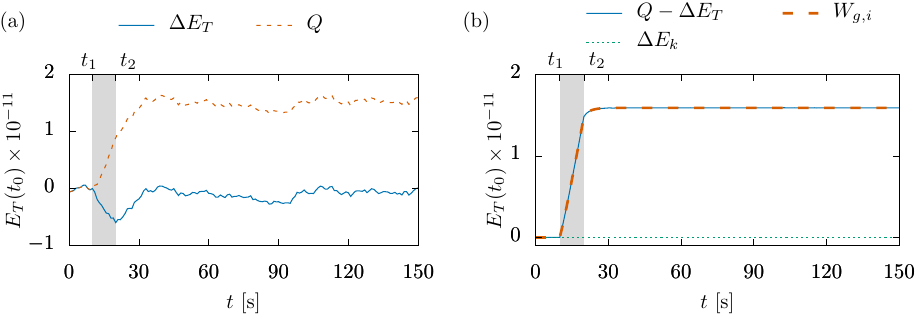}
 \end{center} 
 \caption{The changes in energies, work, and net heat flowing into the system as a function of time in the numerical experiment regarding the quiescent column. Between times $t_1$ and $t_2$, the gravitational field is linearly decreased from $10g$ to $9.9g$. 
 Panel a) shows the change in thermal energy (blue solid line) and the balance between heat flowing in and out of the system (orange dashed line). 
 Panel b) shows the excess heat that left the system (red dashed line), work against gravitational forces (blue dashed line), and change in kinetic energy (green dotted line).}
 \label{fig:figure2a}
\end{figure}

We prepare the system in a stationary state under a gravitational field $10g=98.1$ m/s\textsuperscript{2}.
Next, we decrease the gravitational field by 1\% over 10 s and allow the system to relax to a new stationary state (figure \ref{fig:figure2a}).
We observe that the heat entering the system
\begin{equation}\label{eq:flux}
 Q = \int_{t_1}^{t} \dot{Q} \dd t,
 \qquad
 \dot{Q} =  - \frac{k}{\rho_0 c_v} \left( \int_{\text{top wall}} + \int_{\text{bottom wall}}\right) \dd A \mathbf{n} \cdot  \nabla T.
\end{equation}
During the change of the gravitational field is on the order of $10^{-11}E_T$. 
Due to the isothermal condition $\Delta E_T = 0$ this heat equals the work performed by the gas against the gravitational field.
We observe that simulated quantities do satisfy (\ref{eq:consSmallWithNoEk}) very well. Thus, during the simulation, the following relation holds:
\begin{equation}
 \underbrace{\dd E_T }_{\dd t \int \dd V \frac{\partial (\rho u)}{\partial t} }
 = 
 \underbrace{-W_{g,i}}_{\dd t  \int \dd V \rho \mathbf{v} \cdot \mathbf{g}} 
 + 
 \underbrace{Q}_{\dd t \int \dd V k \partial_z^2 T },
\end{equation}
showing that we indeed simulated the quasi-static process.

\bibliography{thermoBib}

@book{deGroot2013,
  title={Non-equilibrium thermodynamics},
  author={De Groot, S. R. and Mazur, P.},
  year={1984},
  publisher={Dover Publications},
  address={New York}
}

@book{Holyst2012,
  title={Thermodynamics for chemists, physicists and engineers},
  author={Ho{\l}yst, R. and Poniewierski, A.},
  pages={11},
  year={2012},
  publisher={Springer Netherlands},
  address={Dordrecht}
}

@article{Weller1998,
  title={A tensorial approach to computational continuum mechanics using object-oriented techniques},
  author={Weller, H. G. and Tabor, G. and Jasak, H. and Fureby, C.},
  journal={Computational Physics},
  volume={12},
  number={6},
  pages={620--631},
  year={1998},
  publisher={AIP}
}

@inproceedings{Jasak2007,
  title={{OpenFOAM: A C++} library for complex physics simulations},
  author={Jasak, H. and Jemcov, A. and Tukovic, Z.},
  booktitle={International workshop on coupled methods in numerical dynamics},
  volume={1000},
  pages={1--20},
  year={2007},
  address={Zagreb},
  organization={IUC Dubrovnik Croatia}
}

@book{Chandrasekhar2013,
  title={Hydrodynamic and hydromagnetic stability},
  author={Chandrasekhar, S.},
  year={2013},
  publisher={Courier Corporation}
}

@book{Mizerski2021,
  title={Foundations of Convection With Density Stratification},
  author={Mizerski, K. A.},
  year={2021},
  publisher={Springer}
}

@Article{Holyst2023gravity,
AUTHOR = {Ho{\l}yst, R. and {\.Z}uk, P. J. and Makuch, K. and Macio{\l}ek, A. and Giżyński, K.},
TITLE = {Fundamental Relation for the Ideal Gas in the Gravitational Field and Heat Flow},
JOURNAL = {Entropy},
VOLUME = {25},
YEAR = {2023},
NUMBER = {11},
ARTICLE-NUMBER = {1483},
URL = {https://www.mdpi.com/1099-4300/25/11/1483},
PubMedID = {37998175},
ISSN = {1099-4300},
DOI = {10.3390/e25111483}
}

@article{Makuch2024,
  title={Tailoring the first law of thermodynamics for convective flows},
  author={Makuch, K.},
  journal={Physics of Fluids},
  volume={36},
  number={11},
  year={2024},
  publisher={AIP Publishing}
}

@article{Bhatia2012,
  title={The {H}elmholtz-{H}odge decomposition—a survey},
  author={Bhatia, H. and Norgard, G. and Pascucci, V. and Bremer, P.-T.},
  journal={IEEE Transactions on visualization and computer graphics},
  volume={19},
  number={8},
  pages={1386--1404},
  year={2012},
  publisher={IEEE}
}

@misc{githubPageSolver,
      author={{\.Z}uk, P. J.},
      year={2025},
      note = {\url{https://doi.org/10.5281/zenodo.17267833}, \url{https://github.com/pjzuk/psiRayleighBenard}}
}

@article{Holyst2023fundamental,
  title={Fundamental relation for gas of interacting particles in a heat flow},
  author={Ho{\l}yst, R. and Makuch, K. and Gi{\.z}y{\'n}ski, K. and Macio{\l}ek, A. and {\.Z}uk, P. J.},
  journal={Entropy},
  volume={25},
  number={9},
  pages={1295},
  year={2023},
  publisher={MDPI}
}

@article{oberbeck1879warmeleitung,
  title={{\"U}ber die W{\"a}rmeleitung der Fl{\"u}ssigkeiten bei Ber{\"u}cksichtigung der Str{\"o}mungen infolge von Temperaturdifferenzen},
  author={Oberbeck, A.},
  journal={Annalen der Physik},
  volume={243},
  number={6},
  pages={271--292},
  year={1879},
  publisher={Wiley Online Library}
}

@article{lorenz1881leitungsvermogen,
  title={{\"U}ber das Leitungsverm{\"o}gen der Metalle f{\"u}r W{\"a}rme und Elektrizit{\"a}t},
  author={Lorenz, L.},
  journal={Annalen der Physik},
  volume={13},
  pages={422--447},
  year={1881}
}

@article{boussinesq1903theorie,
  title={Theorie analytique de la chaleur vol 2 (Paris: Gauthier-Villars)},
  author={Boussinesq, J.},
  journal={Buoyancy Effects in Fluids},
  year={1903},
  publisher={Cambridge University Press Cambridge, UK}
}

@article{thomson1882changing,
  title={On a changing tesselated structure in certain liquids},
  author={Thomson, J.},
  journal={Proceedings of the Philosophical Society of Glasgow},
  volume={13},
  pages={464--468},
  year={1882}
}

@article{rayleigh1916lix,
  title={{LIX}. {O}n convection currents in a horizontal layer of fluid, when the higher temperature is on the under side},
  author={{R}ayleigh, Lord},
  journal={The London, Edinburgh, and Dublin Philosophical Magazine and Journal of Science},
  volume={32},
  number={192},
  pages={529--546},
  year={1916},
  publisher={Taylor \& Francis}
}

@article{benard1900tourbillons,
  title={Les tourbillons cellulaires dans une nappe liquide},
  author={{B}{\'e}nard, H.},
  journal={Revue G{\'e}n{\'e}rale des Sciences Pures et Appliqu{\'e}es},
  volume={11},
  pages={1261--1271},
  year={1900}
}

@article{Ahlers2009,
  title = {Heat transfer and large scale dynamics in turbulent {R}ayleigh-{B}\'enard convection},
  author = {Ahlers, G. and Grossmann, S. and Lohse, D.},
  journal = {Reviews of Modern Physics},
  volume = {81},
  issue = {2},
  pages = {503--537},
  numpages = {0},
  year = {2009},
  month = {Apr},
  publisher = {American Physical Society},
  doi = {10.1103/RevModPhys.81.503},
  url = {https://link.aps.org/doi/10.1103/RevModPhys.81.503}
}

@article{Bodenschatz2000,
   author = "Bodenschatz, E. and Pesch, W. and Ahlers, G.",
   title = "Recent Developments in {R}ayleigh-{B}{\'e}nard Convection",
   journal= "Annual Review of Fluid Mechanics",
   year = "2000",
   volume = "32",
   number = "Volume 32, 2000",
   pages = "709-778",
   doi = "https://doi.org/10.1146/annurev.fluid.32.1.709",
   url = "https://www.annualreviews.org/content/journals/10.1146/annurev.fluid.32.1.709",
   publisher = "Annual Reviews",
   issn = "1545-4479",
   type = "Journal Article",
  }

@article{ecke2023turbulent,
  title={Turbulent rotating {R}ayleigh--{B}{\'e}nard convection},
  author={Ecke, R. E. and Shishkina, O.},
  journal={Annual Review of Fluid Mechanics},
  volume={55},
  number={1},
  pages={603--638},
  year={2023},
  publisher={Annual Reviews}
}

@article{cross1993pattern,
  title={Pattern formation outside of equilibrium},
  author={Cross, M. C. and Hohenberg, P. C.},
  journal={Reviews of Modern Physics},
  volume={65},
  number={3},
  pages={851},
  year={1993},
  publisher={APS}
}

@article{holyst2025global,
  title={Global non-equilibrium thermodynamics for steady states like never before},
  author={Ho{\l}yst, R and Gi{\.z}y{\'n}ski, K and Macio{\l}ek, A and Makuch, K and Wr{\'o}bel, J and {\.Z}uk, P. J.},
  journal={Europhysics Letters},
  volume={149},
  number={3},
  pages={30001},
  year={2025},
  publisher={IOP Publishing}
}

@article{shang2005test,
  title={Test of steady-state fluctuation theorem in turbulent {R}ayleigh-{B}{\'e}nard convection},
  author={Shang, X.-D. and Tong, P. and Xia, K.-Q.},
  journal={Physical Review E—Statistical, Nonlinear, and Soft Matter Physics},
  volume={72},
  number={1},
  pages={015301},
  year={2005},
  publisher={APS}
}

@article{kita2006principle,
  title={Principle of maximum entropy applied to {R}ayleigh--{B}{\'e}nard convection},
  author={Kita, T.},
  journal={Journal of the Physical Society of Japan},
  volume={75},
  number={12},
  pages={124005},
  year={2006},
  publisher={The Physical Society of Japan}
}

@article{kita2006entropy,
  title={Entropy in nonequilibrium statistical mechanics},
  author={Kita, T.},
  journal={Journal of the Physical Society of Japan},
  volume={75},
  number={11},
  pages={114005--114005},
  year={2006},
  publisher={The Physical Society of Japan}
}

@article{ban2020thermodynamic,
  title={Thermodynamic analysis of bistability in {R}ayleigh--{B}{\'e}nard convection},
  author={Ban, T.},
  journal={Entropy},
  volume={22},
  number={8},
  pages={800},
  year={2020},
  publisher={MDPI}
}

@article{hughes2013available,
  title={Available potential energy in {R}ayleigh--{B}{\'e}nard convection},
  author={Hughes, G. O. and Gayen, B. and Griffiths, R. W.},
  journal={Journal of Fluid Mechanics},
  volume={729},
  pages={R3},
  year={2013},
  publisher={Cambridge University Press}
}

@article{kerr2001energy,
  title={Energy budget in {R}ayleigh-{B}{\'e}nard convection},
  author={Kerr, R. M.},
  journal={Physical Review Letters},
  volume={87},
  number={24},
  pages={244502},
  year={2001},
  publisher={APS}
}

@article{urban2011efficiency,
  title={Efficiency of heat transfer in turbulent {R}ayleigh-{B}{\'e}nard convection},
  author={Urban, P. and Musilov{\'a}, V. and Skrbek, L.},
  journal={Physical Review Letters},
  volume={107},
  number={1},
  pages={014302},
  year={2011},
  publisher={APS}
}

@article{shishkina2007local,
  title={Local heat fluxes in turbulent {R}ayleigh-{B}{\'e}nard convection},
  author={Shishkina, O. and Wagner, C.},
  journal={Physics of Fluids},
  volume={19},
  number={8},
  year={2007},
  publisher={AIP Publishing}
}

@article{mishra2010energy,
  title={Energy spectra and fluxes for {R}ayleigh-{B}{\'e}nard convection},
  author={Mishra, P. K. and Verma, M. K.},
  journal={Physical Review E—Statistical, Nonlinear, and Soft Matter Physics},
  volume={81},
  number={5},
  pages={056316},
  year={2010},
  publisher={APS}
}

@article{siggia1994high,
  title={High {R}ayleigh number convection},
  author={Siggia, E. D.},
  journal={Annual review of fluid mechanics},
  volume={26},
  number={1},
  pages={137--168},
  year={1994},
  publisher={Annual Reviews 4139 El Camino Way, PO Box 10139, Palo Alto, CA 94303-0139, USA}
}

@article{lohse2010small,
  title={Small-scale properties of turbulent {R}ayleigh-{B}{\'e}nard convection},
  author={Lohse, D. and Xia, K.-Q.},
  journal={Annual Review of Fluid Mechanics},
  volume={42},
  number={1},
  pages={335--364},
  year={2010},
  publisher={Annual Reviews}
}

@article{holyst2024direction,
  title={Direction of spontaneous processes in non-equilibrium systems with movable/permeable internal walls},
  author={Ho{\l}yst, R. and {\.Z}uk, P. J. and Macio{\l}ek, A. and Makuch, K. and Gi{\.z}y{\'n}ski, K.},
  journal={Entropy},
  volume={26},
  number={8},
  pages={713},
  year={2024},
  publisher={MDPI}
}

@misc{robertWWW,
  title     = {Lecture on Non-equilibrium thermodynamics for stationary states},
  author    = {Ho{\l}yst, R.},
  year      = 2024,
  note       = {\url{https://www.youtube.com/watch?v=a78VB3FsL6k}}
}

@article{holyst2008heat,
  title={Heat transfer at the nanoscale: evaporation of nanodroplets},
  author={Ho{\l}yst, R. and Litniewski, M.},
  journal={Physical Review Letters},
  volume={100},
  number={5},
  pages={055701},
  year={2008},
  publisher={APS}
}

@article{babin2005evaporation,
  title={Evaporation of a sub-micrometer droplet},
  author={Babin, V. and Holyst, R.},
  journal={The Journal of Physical Chemistry B},
  volume={109},
  number={22},
  pages={11367--11372},
  year={2005},
  publisher={ACS Publications}
}

@article{maciolek2023parameters,
  title={Parameters of state in the global thermodynamics of binary ideal gas mixtures in a stationary heat flow},
  author={Macio{\l}ek, A. and Ho{\l}yst, R. and Makuch, K. and Gi{\.z}y{\'n}ski, K. and {\.Z}uk, P. J.},
  journal={Entropy},
  volume={25},
  number={11},
  pages={1505},
  year={2023},
  publisher={MDPI}
}

@article{holyst2022thermodynamics,
  title={Thermodynamics of stationary states of the ideal gas in a heat flow},
  author={Ho{\l}yst, R. and Makuch, K. and Macio{\l}ek, A. and {\.Z}uk, P. J.},
  journal={The Journal of Chemical Physics},
  volume={157},
  number={19},
  year={2022},
  publisher={AIP Publishing}
}

@article{makuch2023steady,
  title={Steady-state thermodynamics of a system with heat and mass flow coupling},
  author={Makuch, K. and Ho{\l}yst, R. and Gi{\.z}y{\'n}ski, K. and Macio{\l}ek, A. and {\.Z}uk, P. J.},
  journal={The Journal of Chemical Physics},
  volume={159},
  number={19},
  year={2023},
  publisher={AIP Publishing}
}

@article{boulle2022bifurcation,
  title={Bifurcation analysis of two-dimensional {R}ayleigh-{B}{\'e}nard convection using deflation},
  author={Boull{\'e}, N. and Dallas, V. and Farrell, P. E.},
  journal={Physical Review E},
  volume={105},
  number={5},
  pages={055106},
  year={2022},
  publisher={APS}
}

@article{holyst2025cej,
  title={Global stability of steady states in chemical reactors},
  author={Ho{\l}yst, R. and Wr{\'o}bel, J. and Andryszewski, T. and Gi{\.z}y{\'n}ski, K. and Macio{\l}ek, A. and {\.Z}uk, P. J.},
  journal={Chemical Engineering Journal},
  pages={164076},
  year={2025},
  publisher={Elsevier}
}

@article{gizynski2025,
url = {https://doi.org/10.1515/jnet-2025-0003},
title = {The internal energy as a function of state parameters in steady and unsteady Poiseuille flows},
author = {Konrad Giżyński and Karol Makuch and Jan Paczesny and Paweł Żuk and Anna Maciołek and Robert Hołyst},
pages = {679--693},
volume = {50},
number = {4},
journal = {Journal of Non-Equilibrium Thermodynamics},
doi = {doi:10.1515/jnet-2025-0003},
year = {2025},
lastchecked = {2025-10-02}
}

@article{lamee2015,
title = {Multiple stable solutions in the 2D symmetrical two-sided square lid-driven cavity},
journal = {Computers {\&} Fluids},
volume = {119},
pages = {204-212},
year = {2015},
issn = {0045-7930},
doi = {10.1016/j.compfluid.2015.05.022},
author = {T. Lem{\'e}e and G. Kasperski and G. Labrosse and R. Narayanan},
}

@article{Bars2021, 
title={Zonal jets at the laboratory scale: hysteresis and Rossby waves resonance}, 
volume={910}, 
DOI={10.1017/jfm.2020.1000}, 
journal={Journal of Fluid Mechanics}, 
author={Lemasquerier, D. and Favier, B. and Le Bars, M.}, 
year={2021}, 
pages={A18}
}

@article{venturi2010,
  title={Stochastic bifurcation analysis of Rayleigh--B{\'e}nard convection},
  author={Venturi, Daniele and Wan, Xiaoliang and Karniadakis, George Em},
  journal={Journal of fluid mechanics},
  volume={650},
  pages={391--413},
  year={2010},
  publisher={Cambridge University Press}
}

\end{document}